# Review of the three candidate hohlraums in ICF

Xin Li[1], Changshu Wu[1], Zhensheng Dai[1*], Wudi Zheng[1], Yiqing Zhao[1], Huasen Zhang[1], Jianfa Gu[1], Dongguo Kang[1], Fengjun Ge[1], Peijun Gu[1], and Shiyang Zou[1]

[1]*Institute of Applied Physics and Computational Mathematics, Beijing 100094, China*
*Corresponding author: dai_zhensheng@iapcm.ac.cn

In this paper, we give a review of three hohlraum geometries, including cylindrical, octahedral and six-cylinder-port hohlraums, in inertial confinement fusion (ICF) mainly from theoretical side. Every hohlraum has its own strengths and weaknesses. Although there is a problem of drive asymmetry in the cylindrical hohlraums due to some non-ideal factors, the success of ignition is still possible if more laser energy is available beyond the US National Ignition Facility (NIF) in the future. Octahedral hohlraums can provide the high symmetry flux on capsule. However, octahedral hohlraums suffer from several problems due to the complicated three-dimensional plasma conditions inside. And up to now, there is no one target design with the octahedral hohlraums in which each problem can be solved at the same time. Six-cylinder-port hohlraums combine the merits in theory of both cylindrical and octahedral hohlraums to a certain extent. We introduce a target design with good performance by using the six-cylinder-port hohlraums, in which the key issues of concern, such as laser energy, drive symmetry, and laser plasma interaction (LPI), etc, are all balanced.

## 1 Introduction

The indirect-drive approach to ignition and high gain in ICF involves the use of hohlraums [1,2]. A hohlraum consists of a high Z case with laser entrance holes (LEHs). The laser beams entering the LEHs are efficiently converted into x rays at the beam spots on the hohlraum wall in order to create a uniform X-ray radiation bath. The high-Z cavity (hohlraum) is used to smooth the distribution of radiation seen by a fuel-filled capsule. A one-dimensional spherical implosion optimizes the fuel compression and burn. The cylindrical hohlraums are used most often in inertial fusion studies and are chosen as the ignition hohlraums on NIF [3], which used CH capsules and shaped laser pulses that are 14-22 ns in duration. Typical capsule convergence ratios are 25 to 45, so that drive asymmetry can be no more than 1% [1], a demanding specification on hohlraum design. In cylindrical hohlraums, the Legendre polynomial modes $P_2$ and $P_4$ of the flux on capsule are the main asymmetry modes required to be controlled. In theory, the asymmetry can be tuned to fulfilled the specification of ignition capsule in the cylindrical hohlraums with four laser rings. However, the experiments conducted on NIF show that the LPI than expected increases the difficulty of the asymmetry control. The recent work [4] on NIF observed fusion fuel gains exceed unity by using a high-foot implosion method to reduce instability in the implosion. The hot spot at band-time is still far from a sphere shown by the neutron image, and the analysis showed that the P2 symmetry of capsule is as high as -34%. This result shows that the capsule asymmetry is a serious issue, although significant progress has been made toward the goal of ICF ignition at the National Ignition Facility (NIF) [5].

The flux symmetry is strongly depended on hohlraum geometries and laser beam arrangements [6]. Up to now, various designs with different hohlraum geometries and their special beam arrangements have been proposed and investigated [1-3, 7-13]. Up to now, to our knowledge, it seems certainly possible for spherical hohlraums with 6 LEHs (octahedral hohlraums) [7-10] and hohlraums with 6 cylinder ports [11-13], which include their laser arrangements, to make a breakthrough and solve the flux asymmetry problem in cylinder hohlraums on NIF from theoretical side. The three kinds of hohlraum geometry are reviewed in this paper. Other hohlraums, such as the spherical hohlraums with 4 LEHs [14,15], the rugby hohlraums [16-20], et al., which have their own problems, are not discussed here. The flux symmetry and the hohlraum energetic are the most important issues of hohlraum [1-3]. It is noteworthy that a balance must be strike between the two issues because the available laser energy of facilities in reality is usually limited. This laser constraints sometimes have an impact on the effectiveness of the improvements in asymmetry.

Considering the situation that NIF has not achieved ignition so far, it's important to analyze the strength and weakness of each type hohlraum geometry, compare with each other, and explore what to do next. The paper is organized as follows. In Section 2, a review of the cylinder hohlraums is present. Section 3 give a review of the octahedral hohlraums. Section 4 discusses a ignition hohlraum design with six cylinder ports. Section 5 presents the comparison among the three hohlraum geometries. Section 6 summarizes our discussions and conclusions.

## 2 Review of the cylindrical hohlraums

If the hohlraum radius is extremely larger than the capsule radius, hohlraums are effective at smoothing all but the lowest order asymmetry perturbations [1]. In the cylinder hohlraums, the Legendre polynomial modes $P_2$ and $P_4$ of the flux on capsule are the main asymmetry modes required to be controlled because of left/right symmetry and radiation transport smoothing. Analytical models can be very useful in understanding the symmetry behavior of hohlraums [1,21,22]. It is proved that cylindrical hohlraums with four laser rings (two per side) on the hohlraum wall can provide enough symmetry flux to capsule in ideal conditions. The one closer to the waist plane is called the inner ring, while the one closer to the LEH is called the outer ring. It is convenient to characterize the capsule irradiation by the angles, $\theta_{in}$ and $\theta_{out}$, under which the capsule "sees" these laser rings (Fig. 1). After removing a "closed total hohlraum wall" flux, the hohlraum asymmetry is contributed only by the laser rings and the LEHs. The laser rings provide positive flux and the LEHs provide negative flux. We further use a three-dimensional view factor code, based on Ref. 16, to study the symmetry of the cylindrical hohlraums [22]. The view factor calculations show that for laser rings there are two zero points of $\theta$ (~19⁰ and 52⁰), on which the P4 contribution of laser ring is zero. This conclusion has little to do with the width of laser ring, the hohlraum radius, and the brightness of laser ring. Actually, the two zero points are close to the nodes of the Legendre polynomial $P_4(\cos\theta)$ [2]. The two LEHs of a cylindrical hohlraum are on a sphere with the radius $R_{LEH}$. The radius of capsule is denoted as $R_{capsule}$. The smoothing factor of the P4 asymmetry contributed by the LEHs has a node at $R_{LEH}/R_{capsule}\sim 5$ [1]. So the P4 asymmetry can be eliminated in cylindrical hohlraums by putting the laser rings on the zero points of $\theta$ and the LEHs on the node of P4 smoothing factor. According to the view factor calculations, the outer rings contribute positive P2 flux asymmetry, meanwhile the inner rings and the LEHs contribute negative P2 flux asymmetry. And the P2 contributions of different sources can cancel each other out by adjusting the brightness ratio between the outer rings and the inner rings. This is the asymmetry control scheme in the cylindrical hohlraums.

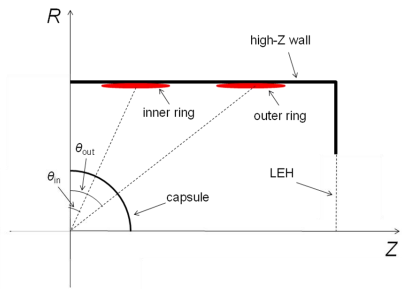

Fig. 1. (color online) Sketch of the cylindrical hohlraum with laser rings

Considering the plasma conditions in hohlraums, the positions of the laser rings time-varies due to the time-dependent motion of wall plasma and the wall albedo also changes over time, which affects the relative brightness between the laser rings and the wall. Considering these effects in the three-dimensional view factor calculations [22], it is proved that the asymmetry control scheme in the cylindrical hohlraums is still adequately effective. Here, we need adjust the relative brightness between the inner rings and the outer rings over time to offset the effect brought by the plasma motion and the time-varying albedo. As a result, the time-dependent P2 is optimized. However, we can only adjust the time-integrated P4 because the P4 asymmetries of the laser rings rely more on the ring positions and it is very difficult to change these positions during the laser duration. Fortunately, simulations show that the P4 flux asymmetry usually changes little during the whole laser duration even after considering the ring motions, so the time-integrated P4 control is enough. The asymmetry control scheme in cylindrical hohlraums has been proved essentially effective by the two-dimensional hydrodynamics simulations in the process of ignition target design [23]. A two dimensional hydrodynamics code, LARED-Integration [24], is used. Here, the relative brightness between the inner rings and the outer rings is adjusted by changing the power ratio between the outer and inner laser cones (tcalled as beam phasing ) on NIF [25]. Fig. 2 shows the second and fourth Legendre coefficients (normalized to the zeroth-order coefficient) of the flux incident on an ignition capsule calculated by using LARED-Integration. The P4 asymmetry remains small during the main pulse, which duration is most crucial for capsule implosion. The $P_2$ and $P_4$ asymmetries can be controlled below 0.5% and 0.1%, respectively, which are far below the typical asymmetry specification [1,3] for ignition.

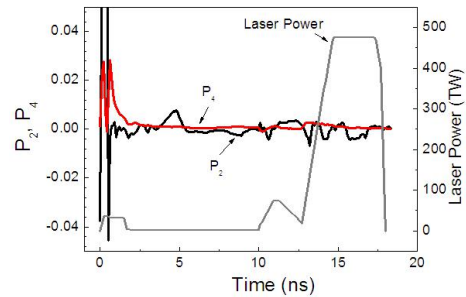

Fig. 2. (color online) Normalized second (black line, left scale) and fourth (red line, left scale) Legendre coefficient of the flux on a typical ignition capsule calculated by LARED-Integration. The laser profile (grey line, right scale).

Of course, the asymmetry control scheme in the cylindrical hohlraums requires that the laser rings do not move too far from the initial positions during the whole laser duration. So the laser transmission to the wall in hohlraums must be unobstructed. There are three factors (Fig. 3) which can destroy the laser transmission in the cylindrical hohlraums. First, the laser energy is absorbed in the radiation-driven high-Z blowoff from the LEHs. The absorption in LEH edges can be avoided by increasing the initial LEH size directly. Second, the laser beam energy is strongly absorbed in high-Z blowoff from the wall (gold bubble) and does not propagate to positions near the hohlraum wall required for the asymmetry control. This problem can be solved by using the initial low-Z fill in cylindrical hohlraums, which is used to block the wall blowoff motion. Third, LPI issues[1] reflect laser energy out of hohlraum directly and reduce the effect of the beam phasing technology. LPI depends on the plasma conditions on the laser paths and the laser conditions strongly, so the hohlraum target should be well designed to avoid LPI in the design stage.

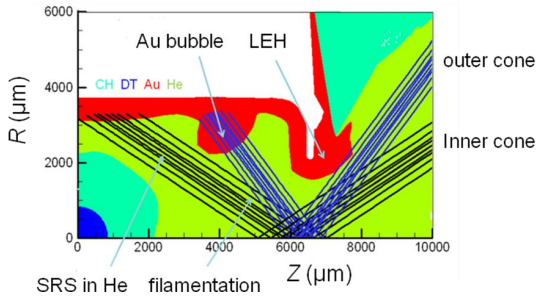

Fig. 3. (color online) The material distribution and the laser paths in a 1/4 typical hohlraum

In the NIF experiments [26], the cylindrical hohlraums can't provide enough high symmetry flux to the ignition capsule when the input laser energy is increased to about 1.9 MJ. This is one of the main reasons that NIF has not achieve ignition yet. About 30%-40% of the energy of the inner cones is reflected off hohlraum directly due to Simulated Raman Scattering (SRS) [26]. As a result, it is very difficult to use beam phasing technology only to control the P2 asymmetry. A crossed-beam energy transfer (CBET) [27] technique is used to transfer a part of laser energy of the outer cones to the inner cones in order to maintain the required symmetry. It makes the asymmetry control on NIF more complicated and the asymmetry swing [26] exists because the amount of the energy transferred varies with time in the main pulse. Ultimately, the SRS on the inner cones is the key factor that results in the asymmetry drive on capsule in NIF cylindrical hohlraums. SRS gain is defined as

$$G_{SRS} \propto \int I_L K_R (n_e/n_c, T_e) dl \quad (1)$$

Where, $I_L$ is laser intensity, $T_e$ and $n_e$ are electron temperature and density respectively, $K_R$ is integral kernel. The integration is along the laser path. The SRS gain is widely used to compare the SRS risk between different target designs [1,28-30]. $K_R$ is strongly dependent on $n_e$ in the parameter space of concern [30]. It is found that the SRS on the inner cones of cylindrical hohlraums is simulated easily within the low-Z plasma deep inside hohlraums [30]. From the theoretical side [30-32], a method of enlarging the size of hohlraum is proposed, which can reduce $n_e$ of the low-Z plasma and decrease the SRS gain. Fig. 4 shows that $n_e/n_c$ in the larger hohlraum is much smaller than that in a smaller hohlraum according the two-dimensional simulations.

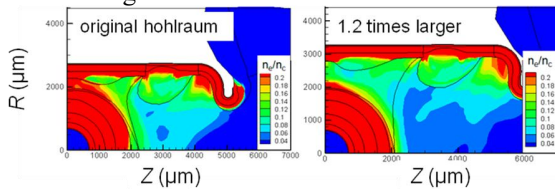

Fig. 4. (color online) Comparison of $n_e/n_c$ between two hohlraums with different sizes.

The experiments [33,34] on NIF have proved the effectiveness of the method. Especially, the shot (N150826) [34] on NIF used a larger hohlraum with the diameter of 6.72 mm to drive a high-foot capsule. The drive duration is about 15 ns long and 0.6 mg/cm³ He is filled into the hohlraum initially. In this shot, about 7 % SRS of the inner cones has been observed, which is much smaller than those in the traditional hohlraums with 5.44 mm or 5.75 mm diameters [30]. And there is very little hot electron due to the small SRS. Of course, larger hohlraums need more laser energy to maintain the drive on capsule. According to the extrapolations based on NIF experiments [35], 2.5 MJ is necessary to achieve the ignition considering other key issues if the method of enlarging hohlraum size is used to control the SRS. Unfortunately, this requirement can't be met because the energy exceeds the maximum of NIF [1] unless NIF is updated or new facilities with more laser energy are built in the future.

Recently, the hohlraums experiment [36] on Shenguang III (SGIII) laser facility [37] and the experiment [38] NIF have proved the effectiveness of the flux asymmetry control scheme conditionally in the cylindrical hohlraums. The yield over clean ratio (YOC) of the target in the experiment on SGIII is designed to be sensitive only to the P₂ flux asymmetry. Fig. 5 shows that the YOC can be as high as 60% once the drive symmetry is adjusted to be good by changing the relative energy ratio to a specific value (~2:1).

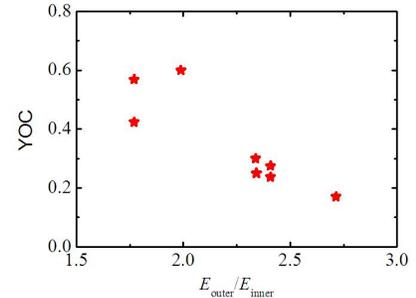

Fig. 5. (color online) YOC varies with the laser energy ratios of the outer and inner cones

NIF utilizes a 2-shock 1 MJ pulse with 340 TW peak power in a near-vacuum Au cylindrical hohlraum and a CH ablator capsule uniformly doped with 1% Si. The purpose of using a near-vacuum hohlraum is to avoid the LPI issues. Similarly, a spherical capsule convergence has been achieved by adjusting the relative powers of the inner and outer cones, which has been demonstrated by the high YOC ratios (about 70%−80%). The asymmetry swing has not been observed and the final P2 shape asymmetry of the hot spot can be adjusted to about -3% at bang time. It means the P2 asymmetry of the flux on capsule is only about 2.3‰ due to the convergence amplification effect [1]. It's noted that there is little LPI effect in the experiment, which makes the asymmetry control easier. Of course, the convergence ratios of capsule in the two experiment are only about 15-20 low and controlling the asymmetry on the higher converge ratios (~30-40) of ignition capsule is still a challenge for the scheme because it needs more shot resources for adjusting the power ratios to obtain more symmetrical fluxes.

## 3 Review of the octahedral hohlraums

The octahedral hohlraums [7-10] (Fig. 6) are proposed to provide the higher flux symmetry to capsule. The asymmetry control scheme is also studied in Ref. [6] in details. The flux asymmetry is contributed only by the laser spots and the LEHs after removing a "closed total hohlraum wall" flux. The six LEHs on a sphere with the octahedral symmetry have no $Y_{lm}$ ($l<4$) asymmetry contribution. Eight laser quads entering each LEH is in one cone at $\theta_L$ as the opening angle that the laser quad beam makes with the LEH normal direction. The laser spots, which equivalent to the LEHs, have also no $Y_{lm}$ ($l<4$) asymmetry contribution. The $Y_{4m}$ asymmetry of the laser spots is proportional to $P_l(\cos 2\theta_L)$. In the octahedral hohlraums [7-10], $\theta_L=55^0$, which is one of the nodes of $P_l(\cos 2\theta_L)$, makes the $Y_{4m}$ asymmetry of the laser spots zero. Because the six LEHs are on a sphere with $R_H/R_C\sim5$, which is very close to the node of the $Y_{4m}$ smooth factor [1], the $Y_{4m}$ asymmetry of the LEHs is also absent. $R_H$ and $R_C$ are the radius of hohlraum and capsule, respectively. Because the smooth factor vanished quickly for large $l$, the $Y_{lm}$ ($l>4$) asymmetry has little influence on the spherical implosion.

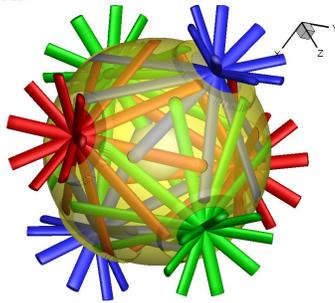

Fig. 6. (color online) Schematic of the octahedral hohlraum with six LEHs and 48 laser quads.

Although the octahedral hohlraums look like perfect in the asymmetry control, there are still several problems in the ignition target designs with the octahedral hohlraums because the balance between the symmetry and other issues, such as the laser energy, LPI, and the plasma motion, et al., must be considered during the ignition target designs. Our studies show that the octahedral hohlraums have five potential problems from theoretical side. Firstly, there is a residual $Y_{4m}$ flux asymmetry of -3‰~-4‰ when $R_H/R_C$ is reduced to about 4 in order to save more laser energy (Fig. 7). In the octahedral hohlraums, $R_H/R_C\sim5$ is required to make the $Y_{4m}$ asymmetry of the LEHs absent. This residual $Y_{4m}$ flux asymmetry is very close to the performance cliff of the typical ignition capsules [3,39] and can't be eliminated by choosing suitable target parameters. Considering the motion of the laser spots due to the plasma blowoff during the laser duration, the equivalent incident angel of the laser beams changes to about $60^0$ in the main pulse, which results in even higher (~ -7‰) $Y_{4m}$ flux asymmetry on capsule.

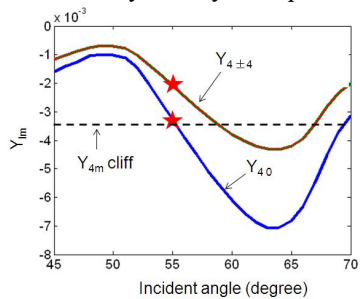

Fig. 7. (color online) Variation of the $Y_{lm}$ asymmetry as the incident angle of laser beams.

Secondly, the 3D laser arrangement [8] shows that some laser beams will be blocked by the high-Z plasma blowoff created by other laser spots which are just above these beams. There is no data to show the affection so the risk is potential. In the view-factor experiment on NIF [40], it is observed that the tip of the Au blowoff driven by laser is 1500 μm far away from the initial position of the Au wall at the end of the laser pulse. The distance between the laser beams and the laser spots above them in the octahedral hohlraums is only about 600 μm (Fig. 8), which results that the Au blowoff goes into the laser path easily and absorbs the laser energy.

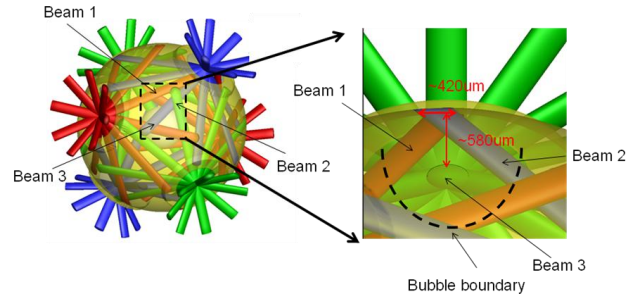

Fig. 8. (color online) Some laser beams can be obsorbed by the Au blowwoff in the octahedral hohlraums.

Thirdly, some laser spots are very close to LEHs since the octahedral hohlraums take a strict injection angel of the laser beams to avoid the laser beam crossing inside hohlraum [8]. Fig. 9 shows the distance between the laser spots and their nearest LEHs is about 100 μm if the nominal spot of 600×400 μm is chosen at best focus. Considering the nominal pointing errors [3] of NIF (each beam is to point within 50 μm of its nominal position, root mean square (RMS) deviation), some laser beams have a chance to transfer outside hohlraum from their neighbor LEHs directly. The problem will be much more serious if the laser focus spots are enlarged to reduce the laser intensities in order to control LPI (see below).

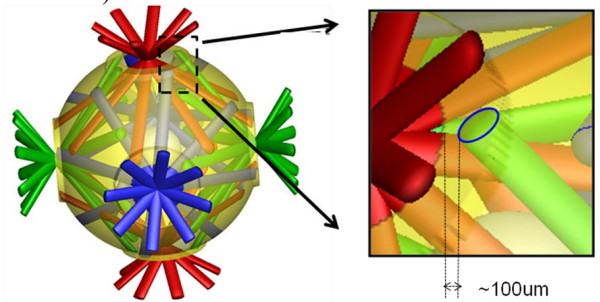

Fig. 9. (color online) Some laser beams can transfer outside their neighborhood LEHs easily.

Fourthly, the high laser intensities [8] (~$3\times10^{15}$ W/cm$^2$) of the octahedral hohlraums can result in high LPI risk since LPI linear gains are proportional to the laser intensity [1]. In the experiments on NIF [41], the Simulated Brillouin Scattering (SBS) is about 10% high on the outer cones of the cylindrical hohlraums with the laser intensities of about $1\times10^{15}$ W/cm$^2$. The intensities in the octahedral hohlraums are about 3 times higher because the laser spot radius must be very small in order to make the beams transfer through the small LEHs successfully with little losses. So the SBS is a big risk in the octahedral hohlraums.

Fifthly, there exists an entering risk of laser beams. Our simulations show that the LEH radius of the octahedral hohlraum [7-10] is still not large enough for laser beams entering successfully because the LEHs are too small ($R_{LEH}$~1 mm) considering the motion of the plasma of the lip of LEH. $R_{LEH}$ is the initial radius of LEH.

Up to now, there is no one ignition target design with the octahedral hohlraums, in which the several problems can be solved at the same time because the target design must balance all key issues impacting the target performance. Of course, the above potential problems come only from the theoretical studies based on the knowledge obtained from the experimental and theoretical studies of the cylindrical hohlraums. We don't know how serious the problems really are because there is little experiments with the real laser arrangement required by the octahedral hohlraums and there is no three-dimensional simulations and no implosion experiments. In the future, more theoretical and experimental studies are still necessary to prove the effectiveness of the octahedral hohlraums used for an ignition capsule in ICF.

## 3 Review of the six-cylinder-port hohlraums

In 2016, we propose a ignition target design with a new hohlraum geometry, the so-called as six-cylinder-port hohlraum (Fig. 10) [11-13]. In this design, a six-cylinder-port hohlraum with 48 laser quads entering in eight LEHs is designed for a typical ignition capsule. Eight laser quads entering each LEH is in one cone at $\theta_L=55^0$ as the opening angle that the laser quad beam makes with the LEH normal direction. The laser power is shaped to meet the requirement of capsule with peak power of 500 TW and total laser energy of 2.3 MJ. The nominal spot of each quad is 600×400 μm at best focus. The LEH size is designed to prevent laser absorption in the plasma of LEH edges. We choose the LEH radius at $R_{LEH}$=1.4 mm, which is much larger than that in the octahedral hohlraums [7-10]. The hohlraum material is gold and the hohlraum is filled with He gas at density 1.5 mg/cc which is confined by a window over the LEH.

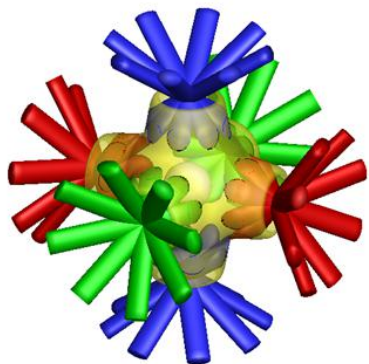

Fig. 10. (color online) Scenography of the hohlraum with 6 cyliner ports, 48 laser quad beams and centrally located fusion capsule.

In the six-cylinder-port hohlraum, the $Y_{lm}$ ($l<4$) asymmetries contributed by the LEHs and the laser spots are all zero because the LEHs and the laser spots maintain the octahedral symmetry, respectively. The problem, that there is a residual $Y_{4m}$ asymmetry contributed by the LEHs in the octahedral hohlraums when $R_H/R_C$ is not on the node of the $Y_{4m}$ smooth factor, is solved in the six-cylinder-port hohlraum. Fig. 10 explains the $Y_{lm}$ asymmetry control scheme. From Eq. (17) in Ref. [6], the 4-order contribution $a_4^{spot,total}$ of laser spots, which is proportion to $P_l(\cos 2\theta_s)$, could be zero at the nodes of $P_l(\cos 2\theta_s)$ at $\theta_s$=15.28$^0$, 35.06$^0$, 54.94$^0$, and 74.72$^0$. $a_l$ is chosen to describe the $l$-order flux asymmetry. $\theta_s$ is the incident angel that the laser beam makes with the LEH normal direction on an outer sphere with radius $R_{outer}$ and $P_l(x)$ is the $l$-order Legendre polynomial function. $a_4^{spot,total}$ is positive when $\theta_s$>74.72$^0$. The same equation also shows that the LEH contribution $a_4^{LEH,total}$<0. The negative sources of LEH and the positive sources of laser spot with $\theta_s$>74.72$^0$ can cancel each other out to make $a_4^{total}$=0 ($a_4^{total}$= $a_4^{spot,total}$+ $a_4^{LEH,total}$). In the six-cylinder-port hohlraum (Fig. 11), the radius $R_{outer,spot}$ of the outer sphere with all laser spots on is smaller than the radius $R_{outer,LEH}$ of the sphere with all LEHs on. So although the laser incident angle $\theta_L$ is 55$^0$, the equivalent laser incident angle $\theta_s$ in sphere with $R_{outer,spot}$ is about 80$^0$. It makes the cancelling possible. Instead, in the octahedral hohlraums[6-8], if $\theta_s$>74.72$^0$, the laser beams will be very close to the hohlraum wall near LEHs and be prone to be absorbed by the wall blowoff plasma near LEHs.

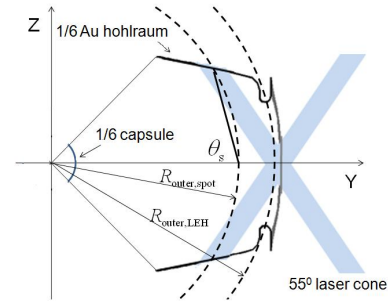

Fig. 11. Scenography of one single port

A three-dimensional view factor code, based on Ref. 13, has proved the effectiveness of the scheme. Although the LEHs with $R_{outer,LEH}/R_C$=4 are not at the node of $Y_{4m}$ smooth factor, the $Y_{4m}$ asymmetry is about zero compared with the octahedral hohlraums. During the design process, we adjust the length of port, the radius of port, the incident angel of laser quad, and the position of laser ring to make the cancelling of the spots and the LEHs happen. Of course, while a large number of target parameters are adjusted to meet the symmetry requirement of capsule, the laser energy, the LPI, and other key issues should be acceptable during the whole adjusting procedure. Time-varying asymmetries are also considered. The $Y_{lm}$ ($l$=1,2,3) asymmetry contributed by the LEHs and the laser spots are all zero during the whole laser duration because the LEHs and the laser spots maintain octahedral symmetry over time, respectively. $Y_{4m}$ is the only remaining asymmetry, which varies with time because the positions of laser spots change over time due to the wall motion. Just like the cylindrical hohlraums [1,3], we can only adjust time-integrated $Y_{4m}$ asymmetry in the six-cylinder-port hohlraum mainly by changing the initial positions of the laser spots. In our target design, the laser spots move to the LEHs about 300 μm during the whole laser duration by analyzing the simulation results (see below). Considering the motion of the laser spots, the three dimensional view factor calculations show that the $Y_{4m}$ asymmetry varies from 1% to 0 (about 1% at the first two nanoseconds and about zero in the main pulse). And the values of the $Y_{4m}$ are much better than the symmetry requirements of the ignition capsule [3,39]. Compared with the six-cylinder-port hohlraums, the $Y_{4m}$ asymmetry during the main pulse in the octahedral hohlraums remains about 3 ‰ -4 ‰ when $R_H/R_C=4$ and the residual asymmetry can't be eliminated by adjusting target parameters.

One of the advantages of the six-cylinder-port hohlraum is that the two-dimensional non-equilibrium radiation hydrodynamics code LARED-Integration can be used to simulate one port because the six ports are relatively independent on each other. It is noticeable that the sum of six single cylinder ports is not equivalent to the hohlraum with six cylinder ports, but the differences are too small to influence the plasma conditions and the energy balance very much. Maps of the spatial distribution of the electron density, the electron temperature, the x-ray emission, and the laser absorption at peak power are shown in Fig. 12. The laser beams in the six-cylinder-port has similar plasma and laser conditions as those on the outer cones of the traditional cylindrical hohlraums on NIF. Since experiment results [26] on NIF do not show obvious LPI on outer cones of the cylindrical hohlraums, there should be little LPI in the laser beams of the six-cylinder-port hohlraum. The laser energy required to drive the hohlraum is estimated as 2.4 MJ, which is very close to the result given by the energy balance [1].

We plan to study the time-varying $Y_{4m}$ asymmetry, the asymmetry of M-band flux and the laser beam entering problem more extensively because there is a larger optimization space in the target design. Of course, in the future, three dimensional simulations are necessary to classify the details of asymmetry, plasma, and efficiency in hohlraums. And a lot of experiments are worthy to be done with the six-cylinder-port hohlraums on the existing facilities.

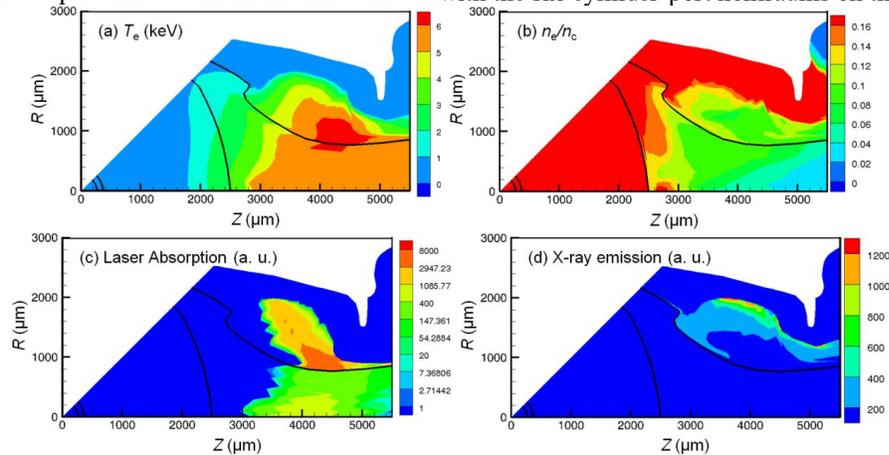

Fig. 12. (color online) Maps of plasma at peak power: (a)$T_e$ in keV. (b) $n_e/n_c$. (c) Laser asorption.(d) x-ray emission

## 4 Comparision among the three hohlraum geometries

Each hohlraum geometry has its own strengths and weaknesses in theoretical side. The six-cylinder-port hohlraums have little SRS risk because they have no laser cones transferring through the low-Z plasma with low electron temperatures and high electron density, which plasma results in high SRS on the inner cones in the cylindrical hohlraums. And the six-cylinder-port hohlraums also have little SBS risk because the laser and plasma conditions are similar to those in the outer cones in the traditional cylindrical hohlraums. But the octahedral hohlraums have higher LPI risk due to the higher laser intensities than those in the cylindrical and six-cylinder-port hohlraums. The $P_2$ asymmetry must be adjusted in the cylindrical hohlraums while the $Y_{lm}$ ($l<4$) asymmetry do not exist in the six-cylinder-port and octahedral hohlraums. In the cylindrical and six-cylinder-port hohlraums, the time-integrated $Y_{4m}$ (or $P_4$) asymmetry can be adjusted by moving the initial positions of the laser spots, while the time-varying $Y_{4m}$ (or $P_4$) asymmetry maintains small values enough for ignition during the whole duration and can be adjusted to zero in the main pulse of laser, on which the symmetry requirement is tightest. However, in the octahedral hohlraums, the $Y_{4m}$ asymmetry can't be eliminated by adjusting target parameters and the values of $Y_{4m}$ are very close to the ignition cliff, which makes the ignition prone to failure. In the octahedral hohlraums, some laser beams can be blocked by the Au blowoff from other laser spots and some laser beams are prone to transfer outside the neighborhood LEHs directly. And the laser beams can be absorbed by the blowoff from the LEHs due to the small LEH radius. The existence of the problems has very small possibility in the cylindrical and six-cylinder-port hohlraums.

The cylindrical hohlraums can be studied with high-fidelity by using two-dimensional non-equilibrium radiation hydrodynamics codes. These codes also can be used to study the plasma conditions, LPI, and hohlraum efficiency in the six-cylinder-port hohlraums not including symmetry. However, it is very difficult for the two-dimensional codes to be used to study those issues in the octahedral hohlraums because the three-dimensional codes are still necessary considering the three-dimensional feature of the complicated plasma conditions inside and applying the three-dimensional codes in studying can go a long way due to the complexity of these codes. The facilities in the world , such as SGIII and NIF, are designed mainly for the cylindrical hohlraums. It is costly to transform the laser arrangements to suit for the six-cylinder-port and octahedral hohlraums. And the transformation is at great risk because the new technical ways, such as the target freezing, radiation flux diagnostics, etc, must be developed to suit for octahedral and six-cylinder-port hohlraums and there are little progress in these respects up to now. The theoretical and experimental studies of the cylindrical hohlraums have a long history. A lot of knowledge and experience about hohlraums are based on these studies. In a word, the technologies in the theoretical and experimental studies of the cylindrical hohlraums are the most mature. There is still a lot of work to do with the octahedral and six-cylinder-port hohlraums in the near future. But the effort will be worth it so long as the ignition with those hohlraums succeed with higher probability in the future.

**5 Conclusion**

In summary, we present a review on the three hohlraum geometries mainly from theoretical side.

It is the major problem in the cylindrical hohlraums that it is difficult to adjust the $P_2$ asymmetry to satisfy the requirement due to the much more SRS on the inner cones than expected although the control scheme is effective. Enlarging the hohlraum size is effective to control the SRS which is proved by the theoretic and experimental studies. But this approach needs more laser energy than that on NIF. The octahedral hohlraums explore a new hohlraum geometry to reduce the flux asymmetry. However, there exist several new problems about the plasma conditions, LPI, laser beam transmission, etc, which results that it is difficult to design a hohlraum for ignition to meet all constrains. On the contrary, although the six-cylinder-port hohlraums also use the octahedral symmetry on the arrangement of the laser spots and the LEHs, the particular design of cylinder port makes the flux asymmetry much smaller than that in the octahedral hohlraums and the existence of the problems, which is probably serious in the octahedral hohlraums, has very small possibility. In this paper, we also present a ignition target design with the six-cylinder-port hohlraums. In this design, the key issues of concern are all balanced and there is no obvious performance flaw.

*Acknowledgments*

This work was supported by the National Natural Science Foundation of China (Grants Nos. 11435011 and 11575034). The authors wish to acknowledge the beneficial discussions with Hao Duan and Yongkun Ding